# High-skilled Human Workers in Non-Routine Jobs are Susceptible to AI Automation but Wage Benefits Differ between Occupations

*Pelin Özgul[1], Marie-Christine Fregin[1], Michael Stops[2], Simon Janssen[2], Mark Levels[1,3]*

**Affiliations:**

1. Maastricht University

2. IAB

3. Wissenschaftszentrum für Sozialforschung Berlin

**Abstract:** Artificial Intelligence (AI) will change human work by taking over specific job tasks, but there is a debate which tasks are susceptible to automation, and whether AI will augment or replace workers and affect wages. By combining data on job tasks with a measure of AI susceptibility, we show that more highly skilled workers are more susceptible to AI automation, and that analytical non-routine tasks are at risk to be impacted by AI. Moreover, we observe that wage growth premiums for the lowest and the highest required skill level appear unrelated to AI susceptibility and that workers in occupations with many routine tasks saw higher wage growth if their work was more strongly susceptible to AI. Our findings imply that AI has the potential to affect human workers differently than canonical economic theories about the impact of technology on work these theories predict.

**One Sentence Summary**: Workers and their wages in occupations with non-routine analytic tasks and tasks that require high skill levels are susceptible to AI automation.

**Main text**:

While the number of tasks that learning machines can undertake is rapidly increasing and their diffusion is currently accelerating in many sectors of the economy, the impact these developments will have on human workers is still unclear (1). Both, the number of jobs that are susceptible to AI automation and the way in which learning machines will substitute or augment human workers remain vigorously contested (2,3). The theories most commonly used to predict how technology impacts work and workers arrive at competing hypotheses. First, skills biased technological change theory (SBTC) posits that technological developments can replace workers in simple tasks that do not require high competency in specific skills, but not workers in complex tasks. As such, it predicts that the more skilled workers are, the less likely it is that they will lose their jobs to machines, and the more likely it is that technology complements their skills (4,5). Although not without debate (6), ample empirical scrutiny demonstrates that SBTC is largely consistent with a wide variety of empirical observations, at least to explain past technological revaluations. For example, SBTC explains why the digital revolution of the 1990s increased the labor share and income of high-skilled workers and suppressed wages among low-skilled workers (7,8). The more general assumption that rising education levels can complement technology also credibly explains why earlier industrial revolutions favored more highly skilled workers (9). Due to this impressive empirical validation, SBTC has long been viewed as the "canonical model" for explaining and predicting the impact of technology on labor (9). However, contrary to the predictions of SBTC, during the digitalization wave of the early 2000s, middle-income jobs were declining relative to both low skilled and high skilled jobs in many Western countries (10). To explain this polarization on the labor market, labor economists have posited a competing theory, based on the assumptions that a) automation changes specific tasks rather than whole jobs, and b) that digital technologies can perform routine tasks, but not non-routine tasks (11). This routine biased technical change theory (RBTC) explains job polarization by positing that computerization of job tasks decreases labor market opportunities for workers that are skilled in routine tasks but increases them for workers who are skilled to perform non-routine tasks (12). This theory predicts that technology can increase the demand for workers with skills for non-routine tasks, relative to workers' routine skills competencies. RBTC indeed offers a convincing explanation for the 2000s' computerization wave: it has helped to understand why the employment structure, employment shares for high-paid professionals and managers and low-paid personal service workers increased in Western Europe between 1993 and 2010, while employment shares of manufacturing and routine office workers strongly decreased (13).

SBTC and RBTC theories were very successful in explaining how past technological changes affected jobs and workers. However, the applicability of these canonical models as a framework for understanding how AI and robots will impact workers and work is a matter of important debate (14). If AI affects job tasks differently than the technologies that drove previous industrial revolutions did, these trusted theories may not suffice to explain how it impacts work and workers. As presented in Acemoglu and Restrepo (2019), the implications of technological change on jobs depends on three main mechanisms; the direct substitution of labour in automated tasks (displacement effect), increasing labour demand in non-automated tasks and creation of new tasks which are accompanied by new skill demands (reinstatement effect). The resulting net wage and employment effects will therefore depend on how these countervailing forces interact and whether they create opportunities for improved productivity gains (15). There is, however, still limited understanding on which workers will be the winners and losers from the introduction of AI. For example, one influential model assumes that accurate decision-making is rewarded on the labor market and that more experienced workers are better at making decisions because their past experience helps them to more accurately predict potential consequences of their decisions. This would explain why professions with high decision-making power (e.g. managers, doctors) have experienced strong wage growth in the past decades. It also would imply that AI may substitute human workers in occupational tasks that more strongly rely on experience (16). Following this reasoning, David Autor assumes that AI could actually devalue expertise, which may usher in a "future where humans supply only generic, undifferentiated labor", but could also very well "complement human capabilities and open new frontiers of possibility", making work much more interesting (17). These conjectures imply that the spread of AI technologies may be qualitative different from earlier technological innovation waves.

Understanding how AI automation will shape work and affect workers is therefore not trivial. If machines can learn to perform non-routine tasks and complex, highly skilled work, AI will affect a much larger and more diverse group of workers than many current models predict (18). This may further exacerbate relative wage premiums across occupations with different task requirements and the rising labor-market returns to interpersonal, social and communication skills (19). However, because the development of learning machines is still an ongoing effort, market diffusion and AI adoption are far from complete, and field experimental studies in firms are all but non-existent. Furthermore, whereas there are ample reliable and valid measures used to capture software or robot diffusion (20), comparably high-quality information about the market adoption of AI is not available, with potentially detrimental consequences for policies aimed at mitigating AI risks (21). Thus, the explanatory power of the existing theoretical views on technical change are hard to test empirically. US-based studies show that high-income occupations may be among the most exposed to AI (22), that occupations exposed to AI may experience small wage increases, and that there are no aggregate employment effects at the industry or occupation level (23). For Germany, Grienberger et al. (2024) find that the share of tasks susceptible to automation increased the most in high-skilled expert jobs (24).

A common interpretation of these findings is that the tasks performed in high wage occupations are more often exposed to AI technologies than tasks in other occupations. Most of the evidence for this interpretation comes from analyses of AI impacts on firms, establishments, occupations, industries, local labor markets or combinations, but the evidence on the potential consequences of AI on workers and tasks is scarce. Consequently, not much is known yet about the variation of the exposure within the mentioned categories and how this variation can be explained. Particularly, it is not clear yet whether high wage occupations *in general* or whether *specific tasks bundles* in high wage occupations are largely susceptible to AI. For assessing if AI actually affects human workers in different ways than previous technologies, understanding this heterogeneity is key. Therefore, in this paper we provide empirical evidence on the variation of AI impacts within jobs at the worker level. We perform four analyses. First, we examine heterogeneity in occupations' susceptibility to AI automation. Second, we describe how susceptible occupational tasks are to AI automation, and determine the extent to which job tasks' susceptibility to automation by AI is related to their skill complexity and routineness level. We compare this with tasks' susceptibility to robotization and software adoption. Third, we explore how AI susceptibility relates to wage growth for workers in occupations with different characteristics. There is a broad agreement in the literature that machines will substitute human workers in some tasks, while augmenting them in others (25). How this plays out exactly is still unclear. We explore this by analyzing how AI susceptibility of tasks affects the relationship between task complexity and routineness and workers' employment opportunities and wages. Fourth, we study whether workers are more likely to leave occupations that are more strongly susceptible to AI. These four analyses allow us to assess how the impact of AI on work and workers can best be understood theoretically, and evaluate whether the AI revolution is indeed fundamentally different from earlier technological disruptions.

To perform these analyses, we combine data on workers, job tasks and automation susceptibility to explore how AI may plausibly impact work and workers. More specifically, we analyze work biographies of about 3 million workers in Germany between 2012-2019 (26). These data come from a representative 25 percent Sample of Integrated Employment Biographies (IEB), provided by the German Institute for Employment Research (IAB). The IEB covers the near-universe labor market history of all individuals in Germany, subject to social security contributions, in all firms, occupations and economic sectors, and includes employees, benefit recipients, unemployed job seekers, and participants in active labor market policy programs, but excludes civil servants and self-employed workers (27). The data contains comprehensive information on workers' socio-demographic characteristics such as their wage, employment and working time status, age, gender, education and workplace. The data also contains detailed industry and occupational classifications including the level of skill requirement to perform workers' respective occupation that is coded at the $5^{th}$ digit of the German Classification of Occupations 2010. We complement this dataset with the explicit measure on the occupational task composition of German occupations that formalizes German occupations into five routine-intensive task categories which enables us to simultaneously measure routineness and skills content of occupations (28). To identify which job tasks are susceptible to machine automation, we use a measure that combines information on skills, tasks and job descriptions with AI patent data. Patent data reveals which tasks the patented innovation enables machines to perform. The linkage of this occupation-patent based information to administrative worker level data renders it possible to determine the impacts of the potential exposure to AI innovation on wages and employment in different worker groups. Various such methods have been proposed and empirically validated as a suitable way to understand the potential impact of AI on jobs (29). The exposure measure developed by Webb (30) allows us to identify separately which occupations can be automated by three different types of technologies: robots, software, and AI. Susceptibility is measured as an exposure percentile, indicating each occupation's relative exposure compared to the average exposure of all occupations. A higher score therefore means a higher susceptibility: a score of 100 would mean that all tasks in an occupation would consist of capabilities described in the particular technology patent, and therefore the more likely to have automation potential by that particular technology (31).

## Occupations in all sectors of the labor market are susceptible to AI automation

In Figure 1, we start with showing the distribution of the AI susceptibility percentiles across broad occupational categories in the German labor market. A higher AI susceptibility score relates to higher potential impact of AI on the occupation. While AI exposure is highest for occupations in IT and manufacturing, the exposure percentile is low for occupations in trade, sales, and tourism, and for occupations in business-related services (i.e., business organization, law, and administration). This implies heterogeneities across occupations that may be caused by differences in the task composition.

[Figure 1 about here]

## Higher income occupations are more susceptible to AI automation

Figure 2 shows the distribution of our automation indices over workers' wage deciles. The y-axis represents the susceptibility to exposure, and the x-axis shows the workers' percentile of wage distribution. The blue line shows the values for robot exposure, the red line for software exposure, and the gray line for the exposure to AI.

We observe three regularities: Firstly, in line with SBTC theory, susceptibility to robotization sharply decreases with workers' wages. Secondly, in line with RBTC theory, middle income jobs are most strongly susceptible to software. Thirdly, in contrast to both SBTC and RBTC, but confirming recent empirical observations on AI (32), we find that AI susceptibility increases with wages, indicating that workers in the top income distribution face the highest AI exposure.

[Figure 2 about here]

## Complex and non-routine work is susceptible to AI automation

How can these empirical regularities be explained? To answer this, we first explore the extent to which jobs with different skill levels are susceptible to various forms of technology adoption. SBTC theory predicts that the more highly skilled employees work in an occupation, the less likely they are susceptible to (a) robotization, (b) software, and (c) AI automation. However, the empirical literature suggests that this cannot be corroborated for AI automation (33). We, therefore, would also expect for our data effects of robotization and software in line with the SBTC theory and an opposite effect of AI automation (hypothesis 1). From RBTC theory, we deduce the expectation that the more non-routine tasks an occupation has, the less likely it is that these occupations are susceptible to (a) robotization, (b) software, and (c) AI automation (hypothesis 2). So far, to the best of our knowledge, this hypothesis is empirically not tested yet for AI automation; therefore, we see no reason to modify hypothesis 2 in favor of a varying effect of AI automation.

In Figure 3, we show how occupations' susceptibilities to robotization, software, and AI automation relate to the task complexity of occupations. Therefore, we divide the workers' occupation into four categories of task complexity. Four required skill levels are defined as follows: [a] unskilled and semi-skilled tasks (helper/assistants), [b] skilled tasks (professionals), [c] complex tasks (specialists), and [d] highly complex tasks (experts) (34). The first panel of Figure 3 shows the results for robot exposure, the second for software exposure, and the third one for exposure to AI.

[Figure 3 about here]

The first and the second panel show that the more complex job tasks are, and thus the more highly skilled workers have to be to perform these tasks, the less susceptible they are to robotization and software. However, the third panel shows that the more complex job tasks are, the more susceptible they are to AI automation. More highly complex job tasks require higher skill levels of workers. This means, as expected, that hypothesis 1 holds for robots, software and for AI with an opposite effect. This replicates findings from comparable analyses on US data and may explain why AI susceptibility appears larger in jobs with higher incomes (35). These findings put again into question whether SBTC can contribute to understanding how AI impacts work and workers.

Figure 4 shows how level of routineness of job tasks is related to technology susceptibility. We distinguish between manual and non-manual jobs, and five levels of routineness: [a] analytical non-routine tasks, [b] interactive non-routine tasks, [c] cognitive routine tasks, [d] manual routine tasks, and [e] manual non-routine tasks (36). Analytical non-routine tasks include for example management, planning, supervision, leadership, direction, controlling, monitoring, and analyzing. Interactive non-routine tasks include commercial and counseling activities such as marketing and advertising. Cognitive routine tasks include administrating, testing, inspecting, measuring, monitoring, and running diagnostics. Manual non-routine tasks include refurbishing, providing services, manual therapy, and bespoke crafts. Manual routine tasks include farming, construction tasks, manufacturing, production, harvesting, operating machines, setting up machines, and typesetting. Again, the first panel shows the results for robot exposure, the second for software exposure, and the third for exposure to AI.

We observe various meaningful patterns in the data. The first panel shows that jobs with manual tasks having the largest share on all tasks to be performed reveal the largest susceptibility to robotization. Non-routine manual jobs are about as susceptible to robotization as routine manual job tasks, which is not in line with RBTC theory. The second panel of Figure 3 shows that interactive non-routine jobs are generally less susceptible to software than routine jobs, but jobs with largest shares in manual and analytical non-routine tasks are about as susceptible to software as cognitive routine jobs. The third panel shows that both, routine and non-routine jobs are susceptible to AI automation. Analytical non-routine job tasks are even more strongly susceptible to AI automation than routine tasks. In contrast, manual tasks seem less susceptible to AI automation than non-manual tasks. Thus, hypothesis 2 holds for robotization and software exposure, but not for AI. Similarly, to the SBTC theory, the derived expectations for the impacts of robotization and software from the RBTC theory can be corroborated but not for AI automatization.

[Figure 4 about here]

## Wages grow faster in occupations susceptible to AI

One major question is how AI affects the workers susceptible to it. Does AI augment workers' skills and make them more productive, or substitute them and force them out of their jobs? One way of examining the potential effects on worker productivity is by analyzing the relation between AI susceptibility and wages. Although the relationship between individual productivity and wages is not uncontroversial, we would maintain that the association between wages and productivity is strong and well-established, and assume that wages are (at least partly) an expression of workers' productivity (37). We further assume a positive relationship between AI susceptibility and actual usage of AI technologies in those jobs. There are studies that support our assumption: based on US job ads, they show evidence for a positive correlation of AI skills demand and the same AI exposure indices that we use and another study corroborates those correlation with German job ads and the same indices (38). If AI augments human workers, we should observe that wages grow faster in occupations that are more susceptible to AI automation (hypothesis 3). To test this, in Table 1, we present estimates from a simple regression of AI susceptibility on wage growth of German workers between 2012 and 2019 (39).

The first column of Table 1 shows the most parsimonious model that only controls for the workers' wage in the base line period of 2012. Including the workers' baseline wage takes into account that workers with higher wages in the baseline period mechanically have lower wage growth. Moreover, the baseline wage absorbs potential unobserved and time-constant heterogeneity that might correlate with the workers' wage growth. The coefficient of main interest shows that an increase of one standard deviation in AI exposure is related to a 2.3 percentage point increase of the workers' wage growth throughout the period between 2012 and 2019. This is effect is very similar to previous results from the US.

Column 2 adds observable worker and firm characteristics and column 3 additionally includes AKM effects of the workers' baseline firm. Column 4 for adds a dummy variable indicating whether the workers' occupation is a sales or administrative occupation, and column 5 adds a full set of base line occupation dummies on the two-digit level. The coefficient estimates remain fairly robust to the inclusion of all control variables and suggest that observable characteristics that commonly have a substantial impact on the workers' wage growth barely influence the relationship between AI exposure and workers' wage growths. This means that alternative interpretations, for example, that AI affects wage growth more in higher paying occupations, are less likely correct. We interpret this as support for hypothesis 3.

[Table 1 about here]

## Workers in Jobs Susceptible to AI are Slightly More Likely to Leave their Jobs

One alternative explanation for the positive association between AI susceptibility and wage growth in occupations may be that AI substitutes human workers, and replaces the lowest performing workers first. In addition to augmenting human workers, AI may completely replace them in specific job tasks. If this is the case, it stands to reason that the lowest performing workers get forced out first. This would imply that workers in occupations that are more strongly susceptible to AI would more likely transition from their job, either to other jobs or to unemployment (hypothesis 4).

Table 2 presents estimates of linear probability analyses of the relation between AI susceptibility and the likelihood that full-time workers switch to marginal jobs or become unemployed or inactive. Again, the first specification shows the most parsimonious specification that only accounts for the workers' baseline wage, and the fifth specification shows the most saturated model including all worker and firm characteristics and a full set of occupation dummies.

Although the estimates suggest that between 2012 and 2019, German workers who were susceptible to AI automation were more likely to switch to (poorer) marginal jobs or in unemployment the effects are extremely small and amount to less than a 0.01 percentage point in the most saturated specification. The observations are in line with the substitution hypothesis, but the associations are very weak and potential effects are small.

[Table 2 about here]

### The wage premium of high skills is (still) unrelated to AI

Our observations are mostly consistent with the assumption that AI susceptibility is positively related to workers' productivity, which suggests augmentation effects. SBTC and RBTC offer useful frameworks for understanding potential mechanisms that explain this. SBTC theory assumes that digital technology largely complements human capital (40). In this view, the more highly skilled workers are, the more likely it is that technologies will complement their skills and increase their productivity, which should generally be associated with higher wage growth. In other words, and related to AI, the higher the required skill level of the job of the worker is, the more likely it is that higher susceptibility to AI technology leads to stronger wage growth (hypothesis 5) (41). This SBTC hypothesis seems to hold true in past industrial revolutions.

The case in point: the economics literature has mostly converged to the conclusion that the IT revolution was biased towards higher skilled workers, and that this offers a compelling explanation for the increased wage inequality that characterized Western labor markets during the early 1990s (42). To test this hypothesis, we analyze the extent to which individuals' daily wages are associated with characteristics of occupations and workers. In Table 3 we present the results of the baseline estimates of simple wage regressions (43). The table presents regression results from a model that extends the specification of Table 1 by adding interaction terms between the measure for AI exposure and the four different skill categories of the workers' baseline occupation. Table 3 follows the structure of Table 1 and present the most parsimonious model in column 1 and the most saturated model in column 3. Rows two throughout five show the isolated coefficient estimates of the skill categories with workers in unskilled jobs as the reference category. As we standardize our measure for AI exposure, the coefficient estimates measure the average wage growth effects for workers in occupations with different skill demand and an average level of AI susceptibility. The estimates reveal that workers in jobs with an average AI exposure experience a higher wage growth if their baseline jobs require more skills. The effects range approximately 3 percentage points for workers in occupation that require some skills to approximately 10 percentage points for workers whose occupations require highly complex skills.

Rows six throughout eight show the results for the interaction terms. The estimates of the interaction terms indicate wage growth changes for the different skills categories if AI susceptibility is one standard deviation higher than average. All coefficient estimates are negative, but rather small and sensitive to model specification. Only the coefficient estimate for the skill category of skilled occupations is statistically significant in all models.

So, we draw conclusions cautiously. We infer from our models that AI susceptibility is positively related to wage growth, regardless of the required skill levels of jobs. We do not find that the wage increase associated with AI automation is stronger in jobs with higher required skill levels. Instead, AI susceptibility slightly moderates the association between workers' skill requirements and their wage growth. The estimates are robust to a variety of specifications, including susceptibility to robotization and software, and the inclusion of various other worker and firm level control variables and region, industry, and occupation fixed effects (44).

[Table 3 about here]

## AI-related wage growth is strongest in jobs with cognitive routine tasks

RBTC assumes that workers in jobs with more abstract tasks should experience a notable increase in productivity from software use, because information technology complements them and allows them to spend more time on abstract tasks in which they specialize. Software use should lead wage increases of workers in abstract task-intensive occupations, because they "benefit from information technology via a virtuous combination of strong complementarities between routine and abstract tasks, elastic demand for services provided by abstract task-intensive occupations, and inelastic labor supply to these occupations over the short and medium term" (45). For workers in manual non-routine jobs, such synergies should – generally – not play a role. If anything, they would more likely replaced by technology. So, the RBTC theory predicts that the more non-routine, non-manual tasks a job has, the more likely it is that higher susceptibility to AI technology would be associated with stronger wage growth in these jobs (hypothesis 6).

[Table 4 about here]

To test this, in Table 4, we interact the measure of AI susceptibility with variables that measure routineness of job tasks. Otherwise, Table 4 follows the same structure as our previous tables. We see that AI susceptibility is positively related to wage growth across the board. Of workers that predominantly perform cognitive routine tasks, those that were in occupations that are susceptible to AI experienced higher wage growth than those that are not ($\delta=.027$). Workers that predominantly performed analytical non-routine tasks, also saw faster wage growth if their occupation was susceptible to AI automation, but the difference is less pronounced ($\delta=027-.012=.015$). This is comparable in size to the AI-related wage growth difference for workers in occupations that have many interactive non-routine tasks ($\delta=.027 -.013=.014$). Interestingly, AI susceptibility appears to ameliorate the lower wages of workers that predominantly perform routine and non-routine manual work. In these occupations, wage growth associated with AI is slightly higher than in non-manual occupations. The estimates for interactive non-routine and manual tasks are sensitive to model specification, which requires caution interpretation. However, we can safely observe that they are not consistent with hypothesis 6, which we based on RBTC theory.

## Interpretations and conclusions

In this paper we explored how AI may plausibly impact work and workers. Models based on RBTC and SBTC theories predict that AI will affect a segmented part of the labor market, and will do so in a more or less predictable way. Our results put some of the core assumptions of these canonical theories into question. SBTC assumes that higher skilled workers are less susceptible to automation, and are mostly augmented by technological innovations, while lower skilled workers are mostly replaced. Our results suggest that the higher skilled are more susceptible to AI automation, but do not become more productive. Rather, AI is positively related to workers' wage growth, regardless of their skills levels. RBTC assumes that routine jobs tasks are most easily automatable, and that workers who perform many non-routine tasks can become more productive, whereas workers with more routine tasks are replaced. To illustrate, Autor, Levy, and Murnane (2003) argue that abstract tasks (that require analytical capabilities, expert competencies, problem-solving capabilities, intuition, creativity) that are common in professional, technical, and managerial occupations, and manual tasks (requiring situational adaptability, visual and language recognition, and in-person interactions) are difficult to computerize (46). Our analyses suggest that manual tasks are indeed relatively low at risk of AI automation, but also that tasks related to management, planning, supervision, leadership, direction, controlling, monitoring, and analyzing are among the most susceptible to automation. We also see that thus far, AI seems to benefit those in cognitive routine tasks most. Interactive non-routine tasks are least susceptible to AI automation, suggesting that tasks that require higher levels of interpersonal- and social skills, in which labor has a comparative advantage, remain as key bottlenecks to the evolution of AI (47). This is in line with optimistic conjectures that AI may serve to augment human labor in many tasks susceptible to AI automation (48). Our observations are also consistent with an alternative explanatory model (49). This model assumes that the extent to which job tasks are susceptible to AI replacement has little to do with task complexity or routineness per se, but is determined by the extent to which necessary preconditions for machine learning are met (e.g. the potential availability of training data, the existence of clearly definable goals and metrics) as well as organizational pre-conditions that allow for successful implementation of AI technologies at workplaces, and limitations that follow from idiosyncratic traits of learning machines (e.g. tasks should have a large tolerance for errors and allow for decision-making to occur in a black box). First empirical tests of this model indeed suggest that current learning machines affect different occupations than earlier automation waves, that most occupations include tasks susceptible to automation, but that only very few occupations are fully automatable (50).

Further research will be necessary to explore the spread of AI technologies in future. According to practical, optimistic views, LLM algorithms already appear capable of mastering a wide array of job tasks, sometimes better than humans (51). However, other studies point to an avenue where, in an open-minded setting, the potential of conceivable AI technologies like LLM must be carefully assessed resulting in a more nuanced pattern of utilization (52).

26  We limit our analysis to the period 2012-2019 to avoid any misallocation due to new classification of German occupations and more importantly, to capture the post-2012 upsurge of AI patenting activities ( see Acemoglu, D., Autor, D., Hazell, J., & Restrepo, P. (2022). Artificial intelligence and jobs: Evidence from online vacancies. Journal of Labor Economics, 40(S1), S293-S340.;). This also ensures that the pandemic-related changes in the labor market from 2020 onwards are excluded from the analysis. As market diffusion of AI in Germany was just starting during this period, any effects we report should be interpreted as lower bound effects.

27  More information about these data is available from the Supplement. Here, we also provide descriptive information and statistics of our measurements.

28  Dengler, K., Matthes, B. & Paulus, W. (2014). Occupational tasks in the German labour market, *FDZ Methodenreport*, 12.

29  Such methods are commonly used, e.g. by; Mann, K., & Püttmann, L. (2018). Benign effects of automation: New evidence from patent texts. Available at SSRN 2959584.; Montobbio, F., Staccioli, J., Virgillito, M. E., & Vivarelli, M. (2021). Labour-saving automation and occupational exposure: a text-similarity measure (No. 2021/43). LEM Working Paper Series ; Grimm, F., & Gathmann, C. (2022). The diffusion of digital technologies and its consequences in the labor market.; Prytkova, E., Petit, F., Li, D., Chaturvedi, S., & Ciarli, T. (2024). The Employment Impact of Emerging Digital Technologies,

30  Webb, M. (2020). *The impact of artificial intelligence on the labor market.* Available at SSRN 3482150

31  However, it is important to note that a score of 100 would still not imply that all activities within that particular occupation can be performed by that technology.

32  Acemoglu, D, Autor, D., Hazell, J., & Restrepo, P. (2022). Artificial intelligence and jobs: Evidence from online vacancies, *Journal of Labor Economics*, 40(S1), S293-S340.

33  Webb, M. (2020). *The impact of artificial intelligence on the labor market.* Available at SSRN 3482150.

Felten, E. W., Raj, M., & Seamans, R. (2019). The occupational impact of artificial intelligence: Labor, skills, and polarization. *NYU Stern School of Business*.

34  Jobs with unskilled or semi-skilled tasks require no formal qualification or only short-term training. Jobs with skilled tasks require a formal vocational education training of at least 2 years. Jobs with complex tasks require a university degree or master craftsman's certificate. Jobs with highly complex tasks require a university degree or similar and, beyond that, profound professional experience or further formal highly specialized qualification certificates like a doctorate or a habilitation. Bundesagentur für Arbeit, Statistik/ Arbeitsmittel .

35  Webb, M. (2020). The impact of artificial intelligence on the labor market. Available at SSRN 3482150.

Acemoglu, D, Autor, D., Hazell, J., & Restrepo, P. (2022). Artificial intelligence and jobs: Evidence from online vacancies, *Journal of Labor Economics*, 40(S1), S293-S340.

36  For more information about our classification of task routineness, see: Dengler, K., Matthes, B. & Paulus, W. (2014). Occupational tasks in the German labour market, FDZ Methodenreport, 12.

37  There is debate about the extent to which wages are a valid and reliable measure of worker productivity, particularly in countries with strong institutional wage setting. The crucial assumption here is that wages are strongly positively related to worker productivity

38  Acemoglu, D, Autor, D., Hazell, J., & Restrepo, P. (2022). Artificial intelligence and jobs: Evidence from online vacancies. *Journal of Labor Economics*, 40(S1), S293-S340.

Peede, L. & Stops, M (2024): Artificial intelligence technologies, skills demand and employment: evidence from linked job ads data. Mimeo.

39  More information about estimation methods can be found in the Supplement

40  Goldin, C. and L.F. Katz. (2008). *The Race Between Education and Technology*. Harvard University Press.

**Funding**:

This project received funding by the Federal Ministry of Labour and Social Affairs based on a decision by the German Bundestag (grant number DKl.00.00018.21; Research Project ai:conomics)

**Author contributions:**

- Conceptualization: ML, MS

- Methodology: ML, PO, MS

- Investigation: ML, PO

- Visualization: PO

- Funding acquisition: MF

- Project administration: MF, MS, SJ

- Supervision: ML, MF, MS, SJ

- Writing – original draft: ML, PO

- Writing – review & editing: ML, MS, MF, PO, SJ

**Competing interests**: Authors declare that they have no competing interests.

**Data and materials availability**: All code is available in the supplementary materials. Data are available from German Institute for Employment Research (IAB). The following restrictions on materials apply: individual-level data must be analyzed on site, after review of research designs and methodology, and is subject to output review to ensure compliance with privacy laws and regulations.

**Fig. 1: Exposure to AI by broad occupational groups**

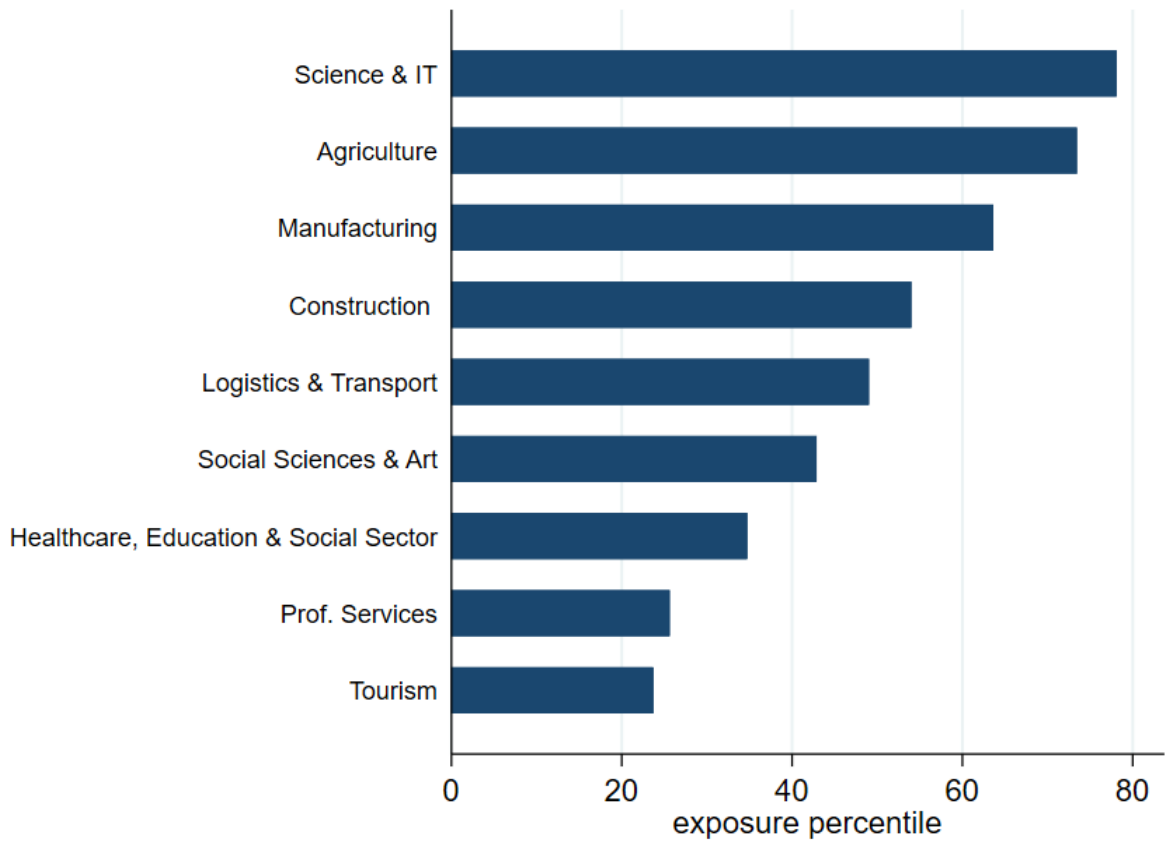

Source: IEB (2012-2019) authors calculations.

**Fig. 2: Susceptibility to automation by wage percentiles**

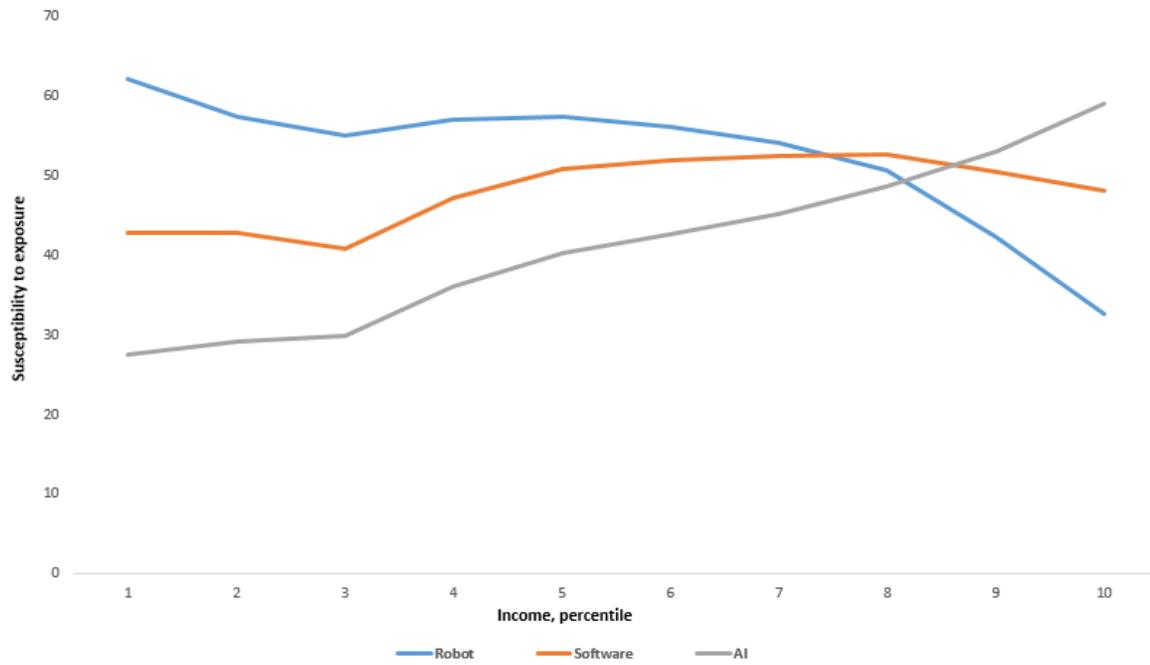

Source: IEB (2012-2019) authors calculations.

**Fig. 3: The relation between task complexity and susceptibility to robots, software and AI automation**

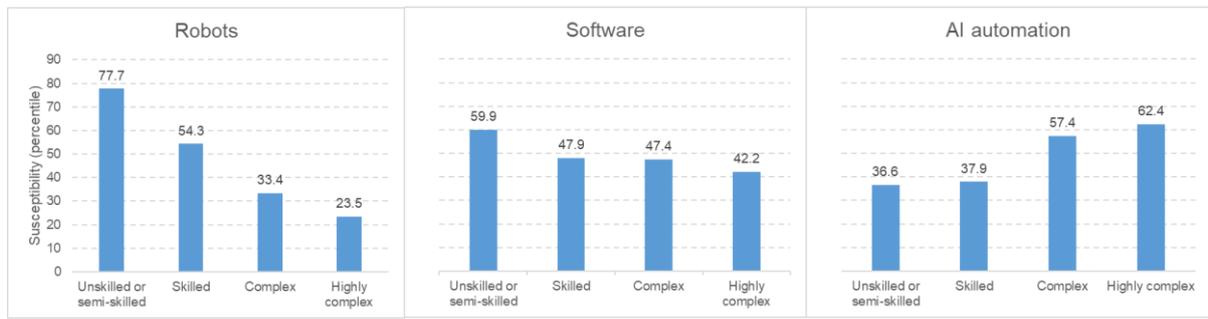

*Source*: IEB (2012-2019), authors' calculations.

**Fig. 4: The relation between task routineness and susceptibility to robots, software, and AI automation, for manual and non-manual work**

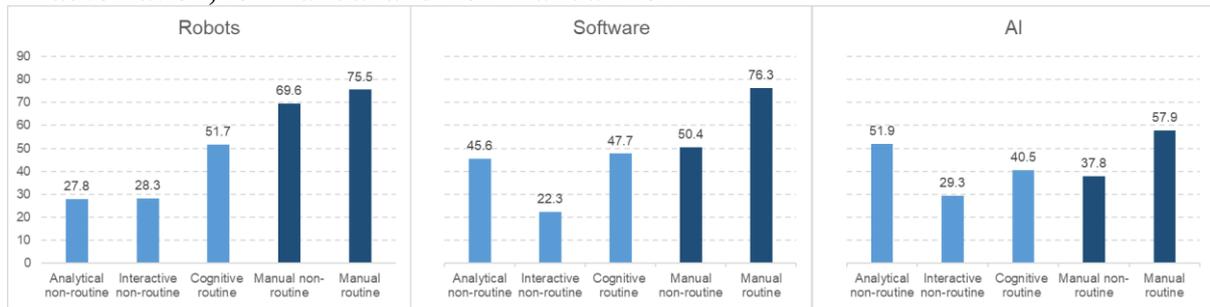

*Source*: IEB (2012-2019), authors' calculations

**Table 1: OLS regression of AI susceptibility on wage growth, 2012-2019**

|  | (1) | (2) | (3) | (4) | (5) |
|---|---|---|---|---|---|
| AI Exposure$_{2012}$ | 0.023*** | 0.011** | 0.018*** | 0.022*** | 0.023*** |
|  | (0.005) | (0.005) | (0.007) | (0.005) | (0.003) |
| Baseline Wage | Yes | Yes | Yes | Yes | Yes |
| Worker Controls | No | Yes | Yes | Yes | Yes |
| Firm Controls | No | Yes | Yes | Yes | Yes |
| AKM Effects | No | No | Yes | Yes | Yes |
| Robot & Software Exposure | No | No | Yes | Yes | Yes |
| Region FEs | No | No | No | Yes | Yes |
| Industry FEs | No | No | No | Yes | Yes |
| Sales & Admin | No | No | No | Yes | No |
| 2-digit Occupation FEs | No | No | No | No | Yes |
| Observations | 3,358,470 | 3,358,428 | 3,221,085 | 3,221,085 | 3,221,085 |
| Adjusted R2 | 0.070 | 0.106 | 0.111 | 0.126 | 0.129 |

*Notes*: This table presents OLS estimates of the relationship between AI susceptibility in 2012 and the difference in log daily wage between 2012-2019. Occupational AI, robot and software exposures are standardized exposure measures over the 5-digit kldb (2010) occupations, based on Webb (2020). The covariates included in the model are reported at the bottom of the table. Column 1 contains only occupational AI exposure controlling for baseline wages in 2012. Columns 2 and 3 include worker and firm controls (education, gender, age, age squared, part-time work, AKM effects and firm size) and robot and software exposure measures. Column 4 include region and industry (3 digit) fixed effects and controls for occupations belonging to either sales or administration. Column 5 includes 2-digit occupation fixed effects. Standard errors within parenthesis are clustered at the 5-digit occupation level.
* $p < 0.05$, ** $p < 0.01$, *** $p < 0.001$.
*Source*: IEB (2012-2019) restricted to full-time workers, liable to social security.

**Table 2: OLS regression of AI susceptibility on employment transitions, 2012-2019**

|  | (1) | (2) | (3) | (4) | (5) |
|---|---|---|---|---|---|
| AI Exposure$_{2012}$ | 0.005*** | 0.001*** | 0.001 | 0.001*** | 0.000** |
|  | (0.001) | (0.000) | (0.000) | (0.000) | (0.000) |
| Worker Controls | No | Yes | Yes | Yes | Yes |
| Firm Controls | No | Yes | Yes | Yes | Yes |
| AKM Effects | No | No | Yes | Yes | Yes |
| Robot & Software Exposure | No | No | Yes | Yes | Yes |
| Region FEs | No | No | No | Yes | Yes |
| Industry FEs | No | No | No | Yes | Yes |
| Sales & Admin | No | No | No | Yes | No |
| 2-digit Occupation FEs | No | No | No | No | Yes |
| Observations | 3,358,470 | 3,358,428 | 3,221,085 | 3,221,085 | 3,221,085 |
| Adjusted R2 | 0.001 | 0.006 | 0.007 | 0.015 | 0.015 |

*Notes*: This table presents OLS estimates of the relationship between AI susceptibility in 2012 and a dummy denoting a change in the employment status from being unemployed in 2012 and being employed in 2019. Occupational AI, robot and software exposures are standardized exposure measures over the 5-digit kldb (2010) occupations, based on Webb (2020). The covariates included in the model are reported at the bottom of the table. Column 1 contains only occupational AI exposure. Columns 2 and 3 include worker and firm controls (education, gender, age, age squared, part-time work, AKM effects and firm size) and robot and software exposure measures. Column 4 include region and industry (3 digit) fixed effects and controls for occupations belonging to either sales or administration. Column 5 includes 2-digit occupation fixed effects. Standard errors within parenthesis are clustered at the 5-digit occupation level.
* $p < 0.05$, ** $p < 0.01$, *** $p < 0.001$.
*Source*: IEB (2012-2019) restricted to full-time workers, liable to social security.

**Table 3: OLS regression of AI susceptibility on wage growth, 2012-2019, for occupations requiring different skill levels**

|  | (1) | (2) | (3) | (4) | (5) |
|---|---|---|---|---|---|
| AI Exposure$_{2012}$ | 0.047*** | 0.030* | 0.048*** | 0.034*** | 0.016* |
|  | (0.016) | (0.017) | (0.017) | (0.011) | (0.009) |
| Unskilled Occupation | ref. | ref. | ref. | ref. | ref. |
| Skilled Occupation | 0.044*** | 0.053*** | 0.032*** | 0.037*** | 0.030*** |
|  | (0.016) | (0.013) | (0.011) | (0.006) | (0.005) |
| Complex Occupation | 0.107*** | 0.117*** | 0.068*** | 0.079*** | 0.068*** |
|  | (0.018) | (0.014) | (0.013) | (0.008) | (0.007) |
| Highly Complex Occupation | 0.165*** | 0.146*** | 0.085*** | 0.103*** | 0.100*** |
|  | (0.035) | (0.021) | (0.024) | (0.017) | (0.014) |
| AI Exposure x Skilled occupation | -0.034** | -0.036** | -0.041*** | -0.024*** | -0.014** |
|  | (0.017) | (0.015) | (0.014) | (0.009) | (0.006) |
| AI Exposure x Complex occupation | -0.034* | -0.029* | -0.036** | -0.018* | -0.006 |
|  | (0.018) | (0.016) | (0.015) | (0.009) | (0.008) |
| AI Exposure x Highly Complex occupation | -0.049 | -0.032 | -0.037* | -0.024* | -0.010 |
|  | (0.033) | (0.021) | (0.020) | (0.014) | (0.011) |
| Baseline Wage | Yes | Yes | Yes | Yes | Yes |
| Worker Controls | No | Yes | Yes | Yes | Yes |
| Firm Controls | No | Yes | Yes | Yes | Yes |
| Task routineness | No | Yes | Yes | Yes | Yes |
| AKM Effects | No | No | Yes | Yes | Yes |
| Robot & Software Exposure | No | No | Yes | Yes | Yes |
| Region FEs | No | No | No | Yes | Yes |
| Industry FEs | No | No | No | Yes | Yes |
| Sales & Admin | No | No | No | Yes | No |
| 2-digit Occupation FEs | No | No | No | No | Yes |
| Observations | 3,358,470 | 3,358,428 | 3,221,085 | 3,221,085 | 3,221,085 |
| Adjusted R2 | 0.078 | 0.113 | 0.114 | 0.129 | 0.131 |

*Notes*: This table presents OLS estimates of the relation between AI exposure and the difference in log daily wage between 2012-2019. Occupational AI, robot and software exposures are standardized exposure measures over the 5-digit kldb (2010) occupations, based on Webb (2020). Occupational skill complexity is an indicator variable based on "requirement level of occupations" for 5-digit kldb (2010) occupations. The covariates included in the model are reported at the bottom of the table. Column 1 contains only occupational AI exposure controlling for baseline wages in 2012. Columns 2 and 3 include worker, occupation and firm controls (education, gender, age, age squared, part-time work, AKM effects, occupational task routiness and firm size) and robot and software exposure measures. Column 4 include region and industry fixed effects (3 digit) and controls for occupations belonging to either sales or administration. Column 5 includes 2-digit occupation fixed effects. Standard errors within parenthesis are clustered at the 5-digit occupation level. * p < 0.05, ** p < 0.01, *** p < 0.001; *Source*: IEB (2012-2019) restricted to full-time workers, liable to social security.

**Table 4: OLS regression of AI susceptibility on wage growth 2012-2019, for occupations requiring different levels of task routineness**

|  | (1) | (2) | (3) | (4) | (5) |
|---|---|---|---|---|---|
| AI Exposure$_{2012}$ | 0.031*** | 0.035*** | 0.045*** | 0.030*** | 0.027** |
|  | (0.006) | (0.013) | (0.013) | (0.011) | (0.011) |
| Cognitive Routine | ref. | ref. | ref. | ref. | ref. |
| Analytical non-routine | 0.031*** | 0.022*** | 0.016** | 0.009** | 0.011** |
|  | (0.009) | (0.007) | (0.007) | (0.004) | (0.005) |
| Analytical non-routine x AI | -0.028*** | -0.022*** | -0.019*** | -0.010*** | -0.012*** |
|  | (0.008) | (0.006) | (0.007) | (0.004) | (0.004) |
| Interactive non-routine | -0.001 | 0.001 | -0.006 | -0.000 | 0.007 |
|  | (0.005) | (0.006) | (0.007) | (0.003) | (0.005) |
| Interactive non-routine x AI | -0.013* | -0.002 | -0.001 | -0.003 | -0.013*** |
|  | (0.007) | (0.006) | (0.006) | (0.003) | (0.004) |
| Manual routine | -0.013** | -0.026*** | -0.024*** | -0.021*** | -0.007* |
|  | (0.005) | (0.005) | (0.005) | (0.003) | (0.004) |
| Manual routine x AI | -0.014** | -0.004 | -0.003 | -0.001 | -0.006*** |
|  | (0.006) | (0.005) | (0.005) | (0.002) | (0.002) |
| Manual non-routine | -0.021*** | -0.023*** | -0.023*** | -0.024*** | -0.011 |
|  | (0.007) | (0.006) | (0.007) | (0.005) | (0.007) |
| Manual non-routine x AI | -0.018** | -0.009 | -0.007 | -0.003 | -0.007* |
|  | (0.008) | (0.006) | (0.007) | (0.004) | (0.004) |
| Baseline Wage | Yes | Yes | Yes | Yes | Yes |
| Worker Controls | No | Yes | Yes | Yes | Yes |
| Firm Controls | No | Yes | Yes | Yes | Yes |
| Occupation skill complexity | No | Yes | Yes | Yes | Yes |
| AKM Effects | No | No | Yes | Yes | Yes |
| Robot & Software Exposure | No | No | Yes | Yes | Yes |
| Region FEs | No | No | No | Yes | Yes |
| Industry FEs | No | No | No | Yes | Yes |
| Sales &Admin | No | No | No | Yes | No |
| 2-digit Occupation FEs | No | No | No | No | Yes |
| Observations | 3,358,470 | 3,358,428 | 3,221,085 | 3,221,085 | 3,221,085 |
| Adjusted R2 | 0.082 | 0.116 | 0.116 | 0.129 | 0.130 |

*Notes*: This table presents OLS estimates of the effects of AI exposure on the difference in log daily wage between 2012-2019. Occupational AI, robot and software exposures are standardized exposure measures over the 5-digit kldb (2010) occupations, based on Webb (2020). Occupational task routine measures are standardized measures over the 3-digit kldb (2010) occupations, based on Dengler et al., (2014). The covariates included in the model are reported at the bottom of the table. Column 1 contains only occupational AI exposure controlling for baseline wages in 2012. Columns 2 and 3 include worker, occupation and firm controls (education, gender, age, age squared, part-time work, AKM effects, occupational skill complexity and firm size) and robot and software exposure measures. Column 4 include region and industry (3 digit) fixed effects and controls for occupations belonging to either sales or administration. Column 5 includes 2-digit occupation fixed effects. Standard errors within parenthesis are clustered at the 5-digit occupation level. * $p < 0.05$, ** $p < 0.01$, *** $p < 0.001$.

*Source*: IEB (2012-2019) restricted to full-time workers, liable to social security & Dengler et al. (2014).

Supplementary Materials for

# High-skilled Human Workers in Non-Routine Jobs are Susceptible to AI Automation but Wage Benefits Differ between Occupations

*Pelin Özgül, Marie-Christine Fregin, Michael Stops, Simon Janssen, Mark Levels*

Corresponding author: m.levels@maastrichtuniversity.nl

**Data and Methods**

Data
  *Occupational level data on exposure to robotics, software, and AI*
  In this section, we give an overview of Webb's (2020) patent-occupation based technology exposure measure. Webb (2020) uses the text of job-task descriptions and the text of patents in a given technological field, to construct a measure of the exposure of tasks to automation[1]. He extracts the verb-noun pairs from each text group and quantifies the overlap between these two texts. Depending on the frequency of overlaps - or in other words, prevalence of such pairs- he assigns a score to the task, and aggregates these task-level scores to the occupation level. With patents acting as indicators of technological progress, and job descriptions acting as indicators of workers' tasks, this overlap eventually indicates how much patenting in a particular technology has been directed at the tasks of any particular occupation (Webb, 2020). Thus, occupations that have a larger fraction of such overlapping tasks are classified as more exposed (Acemoglu et al., 2022). Overall, this method allows him to link a specific technology to specific occupations and to identify which occupations can be automated by a particular technology. He focuses on three different forms of technologies: robotics, software, and AI.

  After providing a numerical (raw) score, he establishes exposure percentiles to indicate each job's relative exposure above or below the average job's exposure to technologies (Webb, 2020). Therefore, the higher the exposure percentile in an occupation, the more likely it is that the occupation consists of capabilities described in the technology patents, and therefore the more likely to have automation potential[2]. We match these occupation-level technology exposure percentiles to register data from Germany that we describe in the next section.

  *Administrative Worker-level Data*
  This section describes the administrative worker data as well as the data preparation procedure to construct a panel data on employment biographies. The main source of German administrative employment records is a 25 % random sample of the Integrated Employment Biographies (IEB) of the IAB[3]. This register data covers the labour market histories of all employees (both regular and irregular[4]), benefit recipients, unemployed job seekers, and participants in active labour market policy measures in Germany with the exception of civil servants and self-employed workers (Genz et al., 2021; Bachmann et al., 2019).

  The construct a panel dataset for our worker-level analysis, we adopt the procedures outlined in Dauth & Eppelsheimer (2020). We restrict our source of data to Employee Histories (BeH). Due to the nonlinearity in employment tracks of individuals (i.e changing jobs, working in multiple jobs, becoming unemployed and etc.) overlapping spells are splitted into episodes such that the resulting data contains parallel spells with same start and end dates. In order to remove all overlapping spells, we select the main episode for workers which is defined by the parallel spell with the longest tenure with a predefined cutoff date (June 30). This allows us to observe each individual only at each point in time and uniquely identified by a worker id and date5.

---

[1] The data on job-tasks descriptions is based on the O*NET database of occupations and task. The patent data is based on Google Patents Public Data, provided by IFI CLAIMS Patent Services (Webb, 2020).
[2] As a word of caution, if an occupation is assigned to a higher score, it means workers in these occupations perform tasks that AI (or robotics / software) could perform, thus they are more exposed to that certain technology. However, a higher exposure score does not necessarily mean that workers in these occupations are in danger of losing work or getting displaced. It is therefore important to note that these exposure scores do not describe the extent to which occupations have already been automated, but how these tasks have the potential to be automated (Webb, 2020).
[3] The IEB covers the universe of all workers in the German labor market that are subject to social security contributions.
[4] Irregular employment includes marginal employment, apprentices, and partial-retirement contract workers (Genz et al. 2021).

Our main outcome variable is the change in real log hourly wages. In the IEB, the workers' daily wages are imputed for those wages that exceed the upper earnings limit for statutory pension insurance and are top-coded at the social security contribution limit (Zimmermann, 2022)6. To recover the missing information on wages, we follow the wage imputation procedure suggested by Card et al. (2013) as outlined in Dauth and Eppelsheimer (2020)7. To make wages comparable across different time periods, we then converted gross daily wages into real wages by using the latest estimates of consumer price index (CPI) from the German Federal Statistical Office which is indexed CPI=100 as of 20158. Lastly, to avoid outliers in the wage distribution, we exclude the top and bottom 1%.

*Data linkage and further steps*

As the Webb (2020) exposure measures are based on US occupations (Standard Occupational Classification, SOC 2010) we rely on the crosswalk provided by Heß et al. (2023) to map Webb's (2020) exposure measures to the German classification of occupations from 2010 at the 5-digit level (Kldb 2010) which provides an up-to-date classification of occupations for Germany (Paulus & Matthes, 2013). The KldB2010 is one of the baseline classifications for the fulfillment of the German employers' obligation to enroll employees with employment information in the social security register.

Once we restrict our sample to all selected employment spells that include June 30th as the cutoff date this leaves us with a representative panel (longitudinal) sample to capture the work biographies of workers over the period 2012-2019.

Finally, we complement the data with establishment-specific wage premia (the so-called AKM effects) for the period of 2010 and 2017 at the establishment level9 and information about the main task type provided by Dengler et al. (2014) at the 3-digit occupation level. Table A.1 provides summary statistics for the estimation sample.

---

[5] See Dauth & Eppelsheimer, 2020 and Antoni et al., 2016 for a detailed explanation of the spell splitting procedure and treatment of overlapping spells.

[6] This means all wage values above the given wage assessment ceiling, which differs based on year and location, are replaced with their given threshold value (Dauth & Eppelsheimer, 2020).

[7] This procedure follows a series of Tobit regressions, controlling for worker characteristics such as year, skill and education groups and regional characteristics.

[8] The data can be accessed [here](here).

[9] See: Lochner, B., Seth, S. & Wolter, S. (2023), AKM effects for German labour market data, FDZ-Methodenreport 01/23, Institute for Employment Research (IAB) for more information on AKM effects.

**Table A.1: Summary statistics along the AI exposure distribution:**

|  | < p25 | | ≥ p75 | |
|---|---|---|---|---|
| Variable | **Mean** | **std. dev** | **Mean** | **std. dev** |
| *Log daily wage* | 4.169 | .961 | 4.637 | .779 |
| *Education* | 4.008 | .964 | | |
|   No Vocational Degree | .177 | .382 | .105 | .307 |
|   Apprenticeship Degree | .68 | .466 | .626 | .484 |
|   University Degree | .143 | .35 | .268 | .443 |
| *Gender* | | | | |
|   Male | .439 | .496 | .771 | .42 |
|   Female | .561 | .496 | .229 | .42 |
| *Age* | 40.79 | 11.88 | 41.50 | 11.24 |
| *Workplace* | | | | |
|   West | .815 | .388 | .833 | .373 |
|   East | .185 | .388 | .167 | .373 |
| *Firm size* | 4.09 | 2.479 | 4.86 | 2.44 |
| *Task Content* | | | | |
|   Analytical non routine | .159 | .366 | .342 | .474 |
|   Interactive non routine | .167 | .373 | .019 | .137 |
|   Cognitive routine | .296 | .457 | .267 | .442 |
|   Manual routine | .106 | .308 | .241 | .428 |
|   Manual non-routine | .271 | .444 | .131 | .338 |
| Observations | 45.164.291 | | 15.444.722 | |

*Notes*: The table reports the sample composition used in wage regressions for bottom and top quartiles of Webb's (2020) AI exposure measure.
*Source*: Webb (2020), IEB (2012-2019), Dengler et al. (2014) & author's calculations

Methods

*Regression models on wage growth 2012-2019*

We estimate the following OLS regression model

$$\Delta y_{2012-2019,i} = \alpha + \beta EXP^{AI}_{O_i} + \boldsymbol{\gamma} \mathbf{x}_i + \boldsymbol{\delta} \mathbf{z}_{f_i} + \theta_{r|f_i} + \vartheta_{s|f_i} + \tau_{SAAD|O_i} + \xi_{2d|O_i} + \varepsilon_i \quad (1)$$

- The term $\Delta y_{2012-2019,i}$ denotes the difference in the logarithm of the daily wage of individual $i$ between 2012 and 2019.
- The term $\alpha$ is the intercept.
- The term $EXP^{AI}_{O_i}$ is the standardized exposure measure for the 5-digit KldB 2010 occupation $o$ of the individual $i$, referring to artificial intelligence. The coefficient $\beta$ measures the change of $\Delta y_{2012-2019,i}$ that corresponds to a change of one standard deviation of the exposure measure $EXP^{AI}_{O\_i}$.

Depending on the specification, further control variables are included:

- The term $\mathbf{x}_i$ denotes the vector with covariates measured at the individual level and $\boldsymbol{\gamma}$ is the vector of the corresponding coefficients. The covariates vector includes (individual) baseline wages in 2012, education, gender, age, age squared, part-time work, the robot and software exposure measures.
- The term $\mathbf{x}_{f_i}$ denotes the vector with covariates measured at the level of the establishment $f$ of individual $i$ and $\boldsymbol{\delta}$ is the vector of the corresponding coefficients. The covariates vector includes, AKM effects and firm size.
- The terms $\theta_{r|f_i}$ denotes fixed effects for region $r$ where the individual's establishment $f$ is located; $\vartheta_{s|f_i}$ denotes a fixed effects of the 3-digit industry $s$ of individual's establishment $f$; $\tau_{SAAD|O_i}$ denotes a dummy for the individual's occupations that belongs to either sales or administration; $\xi_{2d|O_i}$ denotes fixed effects for the 2-digit occupation given the individuals 5-digit occupation.
- The term $\varepsilon_i$ denotes the individual standard errors that are clustered at the 5-digit occupation level.

*Regression models on employment status change*

We estimate the same regression model like in equation (1) with another depending variable:

- The term $\Delta y_{2012-2019,i}$ is a dummy denoting a change in the employment status from being employed in a job subject to social security contributions in full or part time in 2012 and being employed in a marginal job or being unemployed in 2019.
- The coefficient $\beta$ measures the change of the probability to change in a worse job or in unemployment given a change of one standard deviation of the exposure measure $EXP^{C}_{O\_i}$.

*Regression models on wage growth 2012-2019, including required skill levels and interaction terms with the exposure measure*

We estimate the same regression model like in equation (1) and include further terms:

$$\Delta y_{2012-2019,i} = \alpha + \beta EXP^{AI}_{O_i} + \varsigma_{l|O_i} + \psi EXP^{AI}_{O_i} \times \varsigma_{l|O_i} + \boldsymbol{\gamma}\mathbf{x}_i + \boldsymbol{\delta}\mathbf{z}_{f_i} + \theta_{r|f_i} + \vartheta_{s|f_i} + \tau_{SAAD|O_i} + \xi_{2d|O_i} + \varepsilon_i \quad (2)$$

- The term $\varsigma_{l|O_i}$ denotes the required skill level $l$ of the individual's occupation being either "unskilled jobs", "skilled jobs", "complex jobs" or "highly complex jobs".

- The term $EXP_{O_i}^{AI} \times \varsigma_{l|O_i}$ denotes the interaction term of the exposure measure of the individual occupation referring to artificial intelligence and the required skill level.

*Regression models on wage growth 2012-2019, including different levels of task routineness and interaction terms with the exposure measure*

We estimate the same regression model like in equation (2) but differ explaining variables:

- The term $\varsigma_{l|O_i}$ denotes the main tasks type *l* of the individual's occupation being either "manual routine", "cognitive routine", "manual non-routine", "analytical non-routine", or "interactive non-routine".
- The term $EXP_{O_i}^{AI} \times \varsigma_{l|O_i}$ denotes the interaction term of the exposure measure of the individual occupation referring to artificial intelligence and the main tasks type and the corresponding coefficient $\psi$.

**Supplementary Table: Additional analysis and robustness checks**

**Table A.2: OLS regression of AI susceptibility on wage growth, 2012 to 2019**

|  | (1) 2012-2013 | (2) 2012-2014 | (3) 2012-2015 | (4) 2012-2016 | (5) 2012-2017 | (6) 2012-2018 | (7) 2012-2019 |
|---|---|---|---|---|---|---|---|
| AI Exposure$_{2012}$ | 0.010*** | 0.013*** | 0.017*** | 0.019*** | 0.021*** | 0.022*** | 0.023*** |
|  | (0.001) | (0.002) | (0.002) | (0.003) | (0.003) | (0.003) | (0.003) |
| Baseline Wage | Yes | Yes | Yes | Yes | Yes | Yes | Yes |
| Worker Controls | Yes | Yes | Yes | Yes | Yes | Yes | Yes |
| Firm Controls | Yes | Yes | Yes | Yes | Yes | Yes | Yes |
| AKM Effects | Yes | Yes | Yes | Yes | Yes | Yes | Yes |
| Robot & Software Exposure | Yes | Yes | Yes | Yes | Yes | Yes | Yes |
| Region FEs | Yes | Yes | Yes | Yes | Yes | Yes | Yes |
| Industry FEs | Yes | Yes | Yes | Yes | Yes | Yes | Yes |
| 2-digit Occupation FEs | Yes | Yes | Yes | Yes | Yes | Yes | Yes |
| Observations | 3,221,085 | 3,221,085 | 3,221,085 | 3,221,085 | 3,221,085 | 3,221,085 | 3,221,085 |
| Adjusted R2 | 0.029 | 0.045 | 0.069 | 0.090 | 0.106 | 0.119 | 0.129 |

*Notes*: This table presents OLS estimates of the relationship between AI susceptibility in 2012 and the difference in log daily wage between 2012 to 2019. Occupational AI, robot and software exposures are standardized exposure measures over the 5-digit kldb (2010) occupations, based on Webb (2020). The covariates included in the model are reported at the bottom of the table. Standard errors within parenthesis are clustered at the 5-digit occupation level.
* $p < 0.05$, ** $p < 0.01$, *** $p < 0.001$.
*Source*: IEB (2012-2019) restricted to full-time workers, liable to social security.

**Table A.3: OLS regression of AI susceptibility on employment transitions, 2012 to 2019**

|  | (1) 2012-2013 | (2) 2012-2014 | (3) 2012-2015 | (4) 2012-2016 | (5) 2012-2017 | (6) 2012-2018 | (7) 2012-2019 |
|---|---|---|---|---|---|---|---|
| AI Exposure$_{2012}$ | 0.001*** | 0.001** | 0.001*** | 0.001*** | 0.001** | 0.001* | 0.00* |
|  | (0.00) | (0.00) | (0.00) | (0.00) | (0.00) | (0.00) | (0.00) |
| Baseline Wage | Yes | Yes | Yes | Yes | Yes | Yes | Yes |
| Worker Controls | Yes | Yes | Yes | Yes | Yes | Yes | Yes |
| Firm Controls | Yes | Yes | Yes | Yes | Yes | Yes | Yes |
| AKM Effects | Yes | Yes | Yes | Yes | Yes | Yes | Yes |
| Robot & Software Exposure | Yes | Yes | Yes | Yes | Yes | Yes | Yes |
| Region FEs | Yes | Yes | Yes | Yes | Yes | Yes | Yes |
| Industry FEs | Yes | Yes | Yes | Yes | Yes | Yes | Yes |
| 2-digit Occupation FEs | Yes | Yes | Yes | Yes | Yes | Yes | Yes |
| Observations | 3,221,085 | 3,221,085 | 3,221,085 | 3,221,085 | 3,221,085 | 3,221,085 | 3,221,085 |
| Adjusted R2 | 0.020 | 0.024 | 0.022 | 0.018 | 0.015 | 0.015 | 0.015 |

*Notes*: This table presents OLS estimates of the relationship between AI susceptibility in 2012 and a dummy denoting annual change in the employment status from being unemployed in 2012 to being employed in the year following. Occupational AI, robot and software exposures are standardized exposure measures over the 5-digit kldb (2010) occupations, based on Webb (2020). The covariates included in the model are reported at the bottom of the table. Standard errors within parenthesis are clustered at the 5-digit occupation level.
* $p < 0.05$, ** $p < 0.01$, *** $p < 0.001$.
*Source*: IEB (2012-2019) restricted to full-time workers, liable to social security.

**Table A.4: OLS regression of AI susceptibility on wage growth, 2012 to 2019, for occupations requiring different skill levels**

|  | (1) 2012-2013 | (2) 2012-2014 | (3) 2012-2015 | (4) 2012-2016 | (5) 2012-2017 | (6) 2012-2018 | (7) 2012-2019 |
|---|---|---|---|---|---|---|---|
| AI Exposure$_{2012}$ | 0.009*** (0.003) | 0.014*** (0.005) | 0.015*** (0.005) | 0.016** (0.007) | 0.017** (0.008) | 0.017** (0.008) | 0.016* (0.009) |
| Unskilled Occupation | ref. | ref. | ref. | ref. | ref. | ref. | ref. |
| Skilled Occupation | 0.018*** (0.002) | 0.021*** (0.003) | 0.022*** (0.003) | 0.024*** (0.004) | 0.025*** (0.004) | 0.027*** (0.004) | 0.030*** (0.005) |
| Complex Occupation | 0.033*** (0.003) | 0.041*** (0.004) | 0.047*** (0.005) | 0.053*** (0.006) | 0.059*** (0.006) | 0.062*** (0.007) | 0.068*** (0.007) |
| Highly.complex Occupation | 0.048*** (0.006) | 0.062*** (0.008) | 0.073*** (0.010) | 0.081*** (0.012) | 0.089*** (0.013) | 0.094*** (0.014) | 0.100*** (0.014) |
| AI Exposure x Skilled occupation | -0.008*** (0.002) | -0.012*** (0.003) | -0.013*** (0.004) | -0.014*** (0.005) | -0.015*** (0.005) | -0.014** (0.006) | -0.014** (0.006) |
| AI Exposure x Complex occupation | -0.006** (0.003) | -0.009** (0.004) | -0.008 (0.005) | -0.008 (0.006) | -0.008 (0.006) | -0.007 (0.007) | -0.006 (0.008) |
| AI Exposure x Highly Complex occupation | -0.007** (0.004) | -0.011* (0.006) | -0.010 (0.007) | -0.011 (0.008) | -0.011 (0.009) | -0.012 (0.010) | -0.010 (0.011) |
| Baseline Wage | Yes | Yes | Yes | Yes | Yes | Yes | Yes |
| Worker Controls | Yes | Yes | Yes | Yes | Yes | Yes | Yes |
| Firm Controls | Yes | Yes | Yes | Yes | Yes | Yes | Yes |
| occupational task routiness | Yes | Yes | Yes | Yes | Yes | Yes | Yes |
| AKM Effects | Yes | Yes | Yes | Yes | Yes | Yes | Yes |
| Robot & Software Exposure | Yes | Yes | Yes | Yes | Yes | Yes | Yes |
| Region FEs | Yes | Yes | Yes | Yes | Yes | Yes | Yes |
| Industry FEs | Yes | Yes | Yes | Yes | Yes | Yes | Yes |
| 2-digit Occupation FEs | Yes | Yes | Yes | Yes | Yes | Yes | Yes |
| Observations | 3,221,085 | 3,221,085 | 3,221,085 | 3,221,085 | 3,221,085 | 3,221,085 | 3,221,085 |
| Adjusted R2 | 0.030 | 0.046 | 0.071 | 0.091 | 0.107 | 0.120 | 0.131 |

*Notes*: This table presents OLS estimates of the relation between AI exposure and the difference in log daily wage between 2012 to 2019. Occupational AI, robot and software exposures are standardized exposure measures over the 5-digit kldb (2010) occupations, based on Webb (2020). The covariates included in the model are reported at the bottom of the table. Standard errors within parenthesis are clustered at the 5-digit occupation level.
* $p < 0.05$, ** $p < 0.01$, *** $p < 0.001$.
*Source*: IEB (2012-2019) restricted to full-time workers, liable to social security.

**Table A.5: OLS regression of AI susceptibility on wage growth 2012 to 2019, for occupations requiring different levels of task routineness**

|  | (1) 2012-2013 | (2) 2012-2014 | (3) 2012-2015 | (4) 2012-2016 | (5) 2012-2017 | (6) 2012-2018 | (7) 2012-2019 |
|---|---|---|---|---|---|---|---|
| AI Exposure$_{2012}$ | 0.015*** (0.005) | 0.020*** (0.007) | 0.022*** (0.007) | 0.025*** (0.008) | 0.026*** (0.009) | 0.025*** (0.010) | 0.027** (0.011) |
| Cognitive Routine | ref. | ref. | ref. | ref. | ref. | ref. | ref. |
| Analytical non-routine | 0.006*** (0.002) | 0.007** (0.003) | 0.009*** (0.003) | 0.010*** (0.004) | 0.010** (0.004) | 0.011** (0.004) | 0.011** (0.005) |
| Analytical non-routine x AI | -0.004*** (0.001) | -0.006*** (0.002) | -0.009*** (0.002) | -0.010*** (0.003) | -0.011*** (0.003) | -0.012*** (0.003) | -0.012*** (0.004) |
| Interactive non-routine | 0.003 (0.002) | 0.003 (0.003) | 0.004 (0.003) | 0.006 (0.004) | 0.006 (0.005) | 0.005 (0.005) | 0.007 (0.005) |
| Interactive non-routine x AI | -0.003** (0.001) | -0.005*** (0.002) | -0.007*** (0.002) | -0.008*** (0.003) | -0.010*** (0.003) | -0.011*** (0.003) | -0.013*** (0.004) |
| Manual routine | -0.002 (0.002) | -0.003 (0.002) | -0.006** (0.003) | -0.007** (0.003) | -0.007* (0.004) | -0.007* (0.004) | -0.007* (0.004) |
| Manual routine x AI | -0.002** (0.001) | -0.002** (0.001) | -0.003** (0.001) | -0.004** (0.002) | -0.005*** (0.002) | -0.006*** (0.002) | -0.006*** (0.002) |
| Manual non-routine | -0.006** (0.003) | -0.010*** (0.004) | -0.011** (0.005) | -0.013** (0.005) | -0.014** (0.006) | -0.014** (0.006) | -0.011 (0.007) |
| Manual non routine x AI | -0.001 (0.001) | -0.003 (0.002) | -0.005** (0.002) | -0.006** (0.003) | -0.007** (0.003) | -0.007** (0.004) | -0.007* (0.004) |
| Baseline Wage | Yes | Yes | Yes | Yes | Yes | Yes | Yes |
| Worker Controls | Yes | Yes | Yes | Yes | Yes | Yes | Yes |
| Firm Controls occupation skill complexity | Yes | Yes | Yes | Yes | Yes | Yes | Yes |
| AKM Effects | Yes | Yes | Yes | Yes | Yes | Yes | Yes |
| Robot & Software Exposure | Yes | Yes | Yes | Yes | Yes | Yes | Yes |
| Region FEs | Yes | Yes | Yes | Yes | Yes | Yes | Yes |
| Industry FEs | Yes | Yes | Yes | Yes | Yes | Yes | Yes |
| 2-digit Occupation FEs | Yes | Yes | Yes | Yes | Yes | Yes | Yes |
| Observations | 3,221,085 | 3,221,085 | 3,221,085 | 3,221,085 | 3,221,085 | 3,221,085 | 3,221,085 |
| Adjusted R2 | 0.029 | 0.046 | 0.070 | 0.091 | 0.107 | 0.120 | 0.130 |

### Table A.6: OLS regression of Robot susceptibility on wage growth 2012 to 2019

| | (1) 2012-2013 | (2) 2012-2014 | (3) 2012-2015 | (4) 2012-2016 | (5) 2012-2017 | (6) 2012-2018 | (7) 2012-2019 |
|---|---|---|---|---|---|---|---|
| Robot Exposure$_{2012}$ | -0.014*** (0.001) | -0.022*** (0.002) | -0.028*** (0.003) | -0.033*** (0.003) | -0.038*** (0.003) | -0.040*** (0.004) | -0.043*** (0.004) |
| Baseline Wage | Yes | Yes | Yes | Yes | Yes | Yes | Yes |
| Worker Controls | Yes | Yes | Yes | Yes | Yes | Yes | Yes |
| Firm Controls | Yes | Yes | Yes | Yes | Yes | Yes | Yes |
| AKM Effects | Yes | Yes | Yes | Yes | Yes | Yes | Yes |
| Region FEs | Yes | Yes | Yes | Yes | Yes | Yes | Yes |
| Industry FEs | Yes | Yes | Yes | Yes | Yes | Yes | Yes |
| Observations | 3,221,085 | 3,221,085 | 3,221,085 | 3,221,085 | 3,221,085 | 3,221,085 | 3,221,085 |
| Adjusted R2 | .027 | .0425 | .066 | .086 | .102 | .115 | .125 |

### Table A.7: OLS regression of Software susceptibility on wage growth 2012 to 2019

| | (1) 2012-2013 | (2) 2012-2014 | (3) 2012-2015 | (4) 2012-2016 | (5) 2012-2017 | (6) 2012-2018 | (7) 2012-2019 |
|---|---|---|---|---|---|---|---|
| Software Exposure$_{2012}$ | -0.005*** (0.001) | -0.009*** (0.002) | -0.012*** (0.002) | -0.015*** (0.003) | -0.017*** (0.003) | -0.018*** (0.003) | -0.019*** (0.004) |
| Baseline Wage | Yes | Yes | Yes | Yes | Yes | Yes | Yes |
| Worker Controls | Yes | Yes | Yes | Yes | Yes | Yes | Yes |
| Firm Controls | Yes | Yes | Yes | Yes | Yes | Yes | Yes |
| AKM Effects | Yes | Yes | Yes | Yes | Yes | Yes | Yes |
| Region FEs | Yes | Yes | Yes | Yes | Yes | Yes | Yes |
| Industry FEs | Yes | Yes | Yes | Yes | Yes | Yes | Yes |

| | | | | | | |
|---|---|---|---|---|---|---|
| Observations | 3,221,085 | 3,221,085 | 3,221,085 | 3,221,085 | 3,221,085 | 3,221,085 | 3,221,085 |
| Adjusted R2 | .025 | .040 | .063 | .082 | .098 | .110 | .120 |

*Notes*: Occupational AI, robot and software exposures are standardized exposure measures over the 5-digit kldb (2010) occupations, based on Webb (2020). The covariates included in the model are reported at the bottom of the table. Standard errors within parenthesis are clustered at the 5-digit occupation level.
\* $p < 0.05$, \*\* $p < 0.01$, \*\*\* $p < 0.001$.
*Source*: IEB (2012-2019) restricted to full-time workers, liable to social security.

## Table A.8: OLS regression of AI susceptibility on wage growth 2012- 2019, for workers with different education levels

| | (1) | (2) | (3) | (4) | (5) |
|---|---|---|---|---|---|
| AI Exposure$_{2012}$ | 0.022 | 0.008 | 0.011 | 0.010 | 0.002 |
| | (0.014) | (0.010) | (0.012) | (0.012) | (0.008) |
| Low Educ | ref. | ref. | ref. | ref. | ref. |
| Medium Educ | 0.002 | 0.020** | 0.019* | 0.013 | 0.001 |
| | (0.015) | (0.010) | (0.010) | (0.009) | (0.006) |
| High Educ | 0.107*** | 0.084*** | 0.083*** | 0.067*** | 0.054*** |
| | (0.020) | (0.009) | (0.009) | (0.008) | (0.007) |
| Medium Educ x AI | 0.002 | -0.003 | -0.003 | -0.001 | 0.007 |
| | (0.013) | (0.009) | (0.009) | (0.009) | (0.006) |
| High Educ x AI | -0.009 | -0.015 | -0.014 | -0.012 | -0.003 |
| | (0.017) | (0.010) | (0.010) | (0.010) | (0.006) |
| Baseline Wage | Yes | Yes | Yes | Yes | Yes |
| Worker Controls | No | Yes | Yes | Yes | Yes |
| Firm Controls | No | Yes | Yes | Yes | Yes |
| Occupational skill complexity & task Routiness | No | Yes | Yes | Yes | Yes |
| AKM Effects | No | No | Yes | Yes | Yes |
| Robot & Software Exposure | No | No | Yes | Yes | Yes |
| Region FEs | No | No | No | Yes | Yes |
| Industry FEs | No | No | No | Yes | Yes |
| Sales & Admin | No | No | No | Yes | No |
| 2-digit Occupation FEs | No | No | No | No | Yes |
| Observations | 3,358,470 | 3,358,428 | 3,221,085 | 3,221,085 | 3,221,085 |
| Adjusted R2 | 0.076 | 0.115 | 0.115 | 0.119 | 0.124 |

*Notes*: This table presents estimates of the effects of AI exposure on the difference in log daily wage between 2012-2019. Occupational AI, robot and software exposure are standardized exposure measures over the 5-digit kldb (2010) occupations, based on Webb (2020). The covariates included in the model are reported at the bottom of the table. Column 1 contains only occupational AI exposure controlling for baseline wages in 2012. Columns 2 and 3 include worker, occupation and firm controls (gender, age, age squared, part-time work, AKM person-effects, occupational task routiness and firm size) and robot and software exposure measures. Column 4 include region and industry fixed effects and controls for occupations belonging to either sales or administration. Column 5 includes 2-digit occupation fixed effects. Standard errors within parenthesis are clustered at the 5-digit occupation level. * $p < 0.05$, ** $p < 0.01$, *** $p < 0.001$.
*Source*: IEB (2012-2019) restricted to full-time workers, liable to social security.

**Table A.9: OLS regression of AI susceptibility on wage growth, 2012 - 2019, for males**

|  | (1) | (2) | (3) | (4) | (5) |
|---|---|---|---|---|---|
| AI Exposure$_{2012}$ | 0.011 | 0.010* | 0.022*** | 0.026*** | 0.023*** |
|  | (0.007) | (0.005) | (0.007) | (0.006) | (0.004) |
| Baseline Wage | Yes | Yes | Yes | Yes | Yes |
| Worker Controls | No | Yes | Yes | Yes | Yes |
| Firm Controls | No | Yes | Yes | Yes | Yes |
| AKM Effects | No | No | Yes | Yes | Yes |
| Robot & Software Exposure | No | No | Yes | Yes | Yes |
| Region FEs | No | No | No | Yes | Yes |
| Industry FEs | No | No | No | Yes | Yes |
| Sales & Admin | No | No | No | Yes | No |
| 2-digit Occupation FEs | No | No | No | No | Yes |
| Observations | 1,914,312 | 1,914,294 | 1,875,732 | 1,875,732 | 1,875,732 |
| Adjusted R2 | 0.075 | 0.110 | 0.116 | 0.133 | 0.136 |

*Notes*: This table presents OLS estimates of the relationship between AI susceptibility in 2012 and the difference in log daily wage between 2012-2019 for males. Occupational AI, robot and software exposures are standardized exposure measures over the 5-digit kldb (2010) occupations, based on Webb (2020). The covariates included in the model are reported at the bottom of the table. Column 1 contains only occupational AI exposure controlling for baseline wages in 2012. Columns 2 and 3 include worker and firm controls (education, age, age squared, part-time work, AKM effects and firm size) and robot and software exposure measures. Column 4 include region and industry (3 digit) fixed effects and controls for occupations belonging to either sales or administration. Column 5 includes 2-digit occupation fixed effects. Standard errors within parenthesis are clustered at the 5-digit occupation level.
* $p < 0.05$, ** $p < 0.01$, *** $p < 0.001$.
*Source*: IEB (2012-2019) restricted to full-time male workers, liable to social security.

**Table A.10: OLS regression of AI susceptibility on wage growth, 2012 to 2019, for occupations requiring different skill levels, for males**

|  | (1) | (2) | (3) | (4) | (5) |
|---|---|---|---|---|---|
| AI Exposure$_{2012}$ | 0.020 | 0.035** | 0.045*** | 0.029*** | 0.004 |
|  | (0.014) | (0.015) | (0.015) | (0.010) | (0.010) |
| Unskilled Occupation | ref. | ref. | ref. | ref. | ref. |
| Skilled Occupation | 0.039*** | 0.046*** | 0.026** | 0.033*** | 0.029*** |
|  | (0.014) | (0.012) | (0.011) | (0.006) | (0.004) |
| Complex Occupation | 0.133*** | 0.131*** | 0.078*** | 0.086*** | 0.077*** |
|  | (0.013) | (0.013) | (0.016) | (0.010) | (0.008) |
| H.complex Occupation | 0.173*** | 0.147*** | 0.089*** | 0.111*** | 0.112*** |
|  | (0.043) | (0.026) | (0.032) | (0.019) | (0.013) |
| AI Exposure x Skilled occupation | -0.023 | -0.034** | -0.034*** | -0.014** | -0.006 |
|  | (0.017) | (0.013) | (0.013) | (0.007) | (0.005) |
| AI Exposure x Complex occupation | -0.024 | -0.034** | -0.030** | -0.008 | 0.004 |
|  | (0.015) | (0.014) | (0.013) | (0.007) | (0.006) |
| AI Exposure x Highly Complex occupation | -0.019 | -0.027 | -0.025 | -0.014 | -0.001 |
|  | (0.036) | (0.023) | (0.023) | (0.014) | (0.009) |
| Baseline Wage | Yes | Yes | Yes | Yes | Yes |
| Worker Controls | No | Yes | Yes | Yes | Yes |
| Firm Controls | No | Yes | Yes | Yes | Yes |
| occupational task routiness | No | Yes | Yes | Yes | Yes |
| AKM Effects | No | No | Yes | Yes | Yes |
| Robot & Software Exposure | No | No | Yes | Yes | Yes |
| Region FEs | No | No | No | Yes | Yes |
| Industry FEs | No | No | No | Yes | Yes |
| Sales & Admin | No | No | No | Yes | No |
| 2-digit Occupation FEs | No | No | No | No | Yes |
| Observations | 1,914,312 | 1,914,294 | 1,875,732 | 1,875,732 | 1,875,732 |
| Adjusted R2 | 0.087 | 0.118 | 0.119 | 0.135 | 0.138 |

*Notes*: This table presents OLS estimates of the relation between AI exposure and the difference in log daily wage between 2012-2019 for males. Occupational AI, robot and software exposures are standardized exposure measures over the 5-digit kldb (2010) occupations, based on Webb (2020). Occupational skill complexity is an indicator variable based on "requirement level of occupations" for 5-digit kldb (2010) occupations. The covariates included in the model are reported at the bottom of the table. Column 1 contains only occupational AI exposure controlling for baseline wages in 2012. Columns 2 and 3 include worker, occupation and firm controls (education, age, age squared, part-time work, AKM effects, occupational task routiness and firm size) and robot and software exposure measures. Column 4 include region and industry fixed effects (3 digit) and controls for occupations belonging to either sales or administration. Column 5 includes 2-digit occupation fixed effects. Standard errors within parenthesis are clustered at the 5-digit occupation level. * p < 0.05, ** p < 0.01, *** p < 0.001;
*Source*: IEB (2012-2019) restricted to full-time male workers, liable to social security.

## Table A.11: OLS regression of AI susceptibility on wage growth 2012 to 2019, for occupations requiring different levels of task routineness, for males

|  | (1) | (2) | (3) | (4) | (5) |
|---|---|---|---|---|---|
| AI Exposure$_{2012}$ | 0.022*** | 0.009 | 0.024** | 0.008 | 0.005 |
|  | (0.005) | (0.011) | (0.010) | (0.010) | (0.011) |
| Cognitive Routine | ref. | ref. | ref. | ref. | ref. |
| Analytical non-routine | 0.018* | 0.014* | 0.010 | 0.010* | 0.011** |
|  | (0.010) | (0.008) | (0.008) | (0.006) | (0.005) |
| Analytical non-routine x AI | -0.023*** | -0.021*** | -0.017** | -0.012*** | -0.013*** |
|  | (0.009) | (0.006) | (0.007) | (0.004) | (0.004) |
| Interactive non-routine | -0.006 | -0.009** | -0.016*** | -0.002 | -0.002 |
|  | (0.005) | (0.005) | (0.005) | (0.004) | (0.005) |
| Interactive non-routine x AI | -0.009 | -0.004 | -0.002 | -0.002 | -0.010*** |
|  | (0.007) | (0.005) | (0.006) | (0.004) | (0.004) |
| Manual routine | -0.029*** | -0.027*** | -0.027*** | -0.021*** | -0.011** |
|  | (0.006) | (0.005) | (0.005) | (0.003) | (0.005) |
| Manual routine x AI | -0.011* | -0.006 | -0.004 | -0.001 | -0.005* |
|  | (0.006) | (0.004) | (0.004) | (0.003) | (0.002) |
| Manual non-routine | -0.050*** | -0.038*** | -0.038*** | -0.029*** | -0.013 |
|  | (0.008) | (0.006) | (0.006) | (0.005) | (0.008) |
| Manual nonroutine x AI | -0.008 | 0.002 | 0.003 | 0.001 | -0.001 |
|  | (0.007) | (0.006) | (0.006) | (0.003) | (0.003) |
| Baseline Wage | Yes | Yes | Yes | Yes | Yes |
| Worker Controls | No | Yes | Yes | Yes | Yes |
| Firm Controls | No | Yes | Yes | Yes | Yes |
| occupation skill complexity | No | Yes | Yes | Yes | Yes |
| AKM Effects | No | No | Yes | Yes | Yes |
| Robot & Software Exposure | No | No | Yes | Yes | Yes |
| Region FEs | No | No | No | Yes | Yes |
| Industry FEs | No | No | No | Yes | Yes |
| | (1) | (2) | (3) | (4) | (5) |
| Sales &Admin | No | No | No | Yes | No |
| 2-digit Occupation FEs | No | No | No | No | Yes |
| Observations | 1,914,312 | 1,914,294 | 1,875,732 | 1,875,732 | 1,875,732 |
| Adjusted R2 | 0.096 | 0.124 | 0.123 | 0.136 | 0.137 |

*Notes*: This table presents OLS estimates of the effects of AI exposure on the difference in log daily wage between 2012-2019 for males. Occupational AI, robot and software exposures are standardized exposure measures over the 5-digit kldb (2010) occupations, based on Webb (2020). Occupational task routine measures are standardized measures over the 3-digit kldb (2010) occupations, based on Dengler et al., (2014). The covariates included in the model are reported at the bottom of the table. Column 1 contains only occupational AI exposure controlling for baseline wages in 2012. Columns 2 and 3 include worker, occupation and firm controls (education, age, age squared, part-time work, AKM effects, occupational skill complexity and firm size) and robot and software exposure measures. Column 4 include region and industry (3 digit) fixed effects and controls for occupations belonging to either sales or administration. Column 5 includes 2-digit occupation fixed effects. Standard errors within parenthesis are clustered at the 5-digit occupation level. $* p < 0.05$, $** p < 0.01$, $*** p < 0.001$.
*Source*: IEB (2012-2019) restricted to full-time male workers, liable to social security & Dengler et al. (2014).

### Table A.12: OLS regression of AI susceptibility on wage growth, 2012 to 2019, for females

| | (1) | (2) | (3) | (4) | (5) |
|---|---|---|---|---|---|
| AI Exposure$_{2012}$ | 0.020*** | 0.012** | 0.014 | 0.015*** | 0.020*** |
| | (0.006) | (0.006) | (0.009) | (0.005) | (0.005) |
| Baseline Wage | Yes | Yes | Yes | Yes | Yes |
| Worker Controls | No | Yes | Yes | Yes | Yes |
| Firm Controls | No | Yes | Yes | Yes | Yes |
| AKM Effects | No | No | Yes | Yes | Yes |
| Robot & Software Exposure | No | No | Yes | Yes | Yes |
| Region FEs | No | No | No | Yes | Yes |
| Industry FEs | No | No | No | Yes | Yes |
| Sales& Admin | No | No | No | Yes | No |
| 2-digit Occupation FEs | No | No | No | No | Yes |
| Observations | 1,444,158 | 1,444,134 | 1,345,353 | 1,345,353 | 1,345,353 |
| Adjusted R2 | 0.069 | 0.101 | 0.106 | 0.121 | 0.124 |

*Notes*: This table presents OLS estimates of the relationship between AI susceptibility in 2012 and the difference in log daily wage between 2012-2019 for females. Occupational AI, robot and software exposures are standardized exposure measures over the 5-digit kldb (2010) occupations, based on Webb (2020). The covariates included in the model are reported at the bottom of the table. Column 1 contains only occupational AI exposure controlling for baseline wages in 2012. Columns 2 and 3 include worker and firm controls (education, age, age squared, part-time work, AKM effects and firm size) and robot and software exposure measures. Column 4 include region and industry (3 digit) fixed effects and controls for occupations belonging to either sales or administration. Column 5 includes 2-digit occupation fixed effects. Standard errors within parenthesis are clustered at the 5-digit occupation level.
$* p < 0.05$, $** p < 0.01$, $*** p < 0.001$.
*Source*: IEB (2012-2019) restricted to full-time female workers, liable to social security.

**Table A.13: OLS regression of AI susceptibility on wage growth, 2012 to 2019, for occupations requiring different skill levels, for females**

|  | (1) | (2) | (3) | (4) | (5) |
|---|---|---|---|---|---|
| AI Exposure$_{2012}$ | 0.021* | 0.002 | 0.041* | 0.028** | 0.023* |
|  | (0.011) | (0.017) | (0.023) | (0.012) | (0.012) |
| Unskilled Occupation | ref. | ref. | ref. | ref. | ref. |
| Skilled Occupation | 0.074*** | 0.068*** | 0.036** | 0.045*** | 0.033*** |
|  | (0.024) | (0.020) | (0.015) | (0.009) | (0.006) |
| Complex Occupation | 0.121*** | 0.106*** | 0.056*** | 0.075*** | 0.061*** |
|  | (0.023) | (0.019) | (0.019) | (0.010) | (0.008) |
| H.complex Occupation | 0.199*** | 0.145*** | 0.076*** | 0.101*** | 0.095*** |
|  | (0.029) | (0.023) | (0.029) | (0.019) | (0.017) |
| AI Exposure x Skilled occupation | -0.013 | -0.014 | -0.034** | -0.025*** | -0.018* |
|  | (0.013) | (0.015) | (0.016) | (0.010) | (0.010) |
| AI Exposure x Complex occupation | -0.015 | -0.015 | -0.038* | -0.022* | -0.017 |
|  | (0.015) | (0.018) | (0.019) | (0.012) | (0.012) |
| AI Exposure x Highly Complex occupation | -0.041 | -0.016 | -0.034 | -0.021 | -0.014 |
|  | (0.025) | (0.020) | (0.022) | (0.016) | (0.014) |
| Baseline Wage | Yes | Yes | Yes | Yes | Yes |
| Worker Controls | No | Yes | Yes | Yes | Yes |
| Firm Controls | No | Yes | Yes | Yes | Yes |
| occupational task routiness | No | Yes | Yes | Yes | Yes |
| AKM Effects | No | No | Yes | Yes | Yes |
| Robot & Software Exposure | No | No | Yes | Yes | Yes |
| Region FEs | No | No | No | Yes | Yes |
| Industry FEs | No | No | No | Yes | Yes |
| Sales & Admin | No | No | No | Yes | No |
| 2-digit Occupation FEs | No | No | No | No | Yes |
| Observations | 1,444,158 | 1,444,134 | 1,345,353 | 1,345,353 | 1,345,353 |
| Adjusted R2 | 0.077 | 0.107 | 0.109 | 0.123 | 0.126 |

*Notes*: This table presents OLS estimates of the relation between AI exposure and the difference in log daily wage between 2012-2019 for females. Occupational AI, robot and software exposures are standardized exposure measures over the 5-digit kldb (2010) occupations, based on Webb (2020). Occupational skill complexity is an indicator variable based on "requirement level of occupations" for 5-digit kldb (2010) occupations. The covariates included in the model are reported at the bottom of the table. Column 1 contains only occupational AI exposure controlling for baseline wages in 2012. Columns 2 and 3 include worker, occupation and firm controls (education, age, age squared, part-time work, AKM effects, occupational task routiness and firm size) and robot and software exposure measures. Column 4 include region and industry fixed effects (3 digit) and controls for occupations belonging to either sales or administration. Column 5 includes 2-digit occupation fixed effects. Standard errors within parenthesis are clustered at the 5-digit occupation level. * $p < 0.05$, ** $p < 0.01$, *** $p < 0.001$;
*Source*: IEB (2012-2019) restricted to full-time female workers, liable to social security.

**Table A.14: OLS regression of AI susceptibility on wage growth 2012 to 2019, for occupations requiring different levels of task routineness, for females**

| | (1) | (2) | (3) | (4) | (5) |
|---|---|---|---|---|---|
| AI Exposure$_{2012}$ | 0.029*** | 0.058*** | 0.059*** | 0.048*** | 0.047*** |
| | (0.007) | (0.014) | (0.016) | (0.010) | (0.009) |
| Cognitive Routine | ref. | ref. | ref. | ref. | ref. |
| Analytical non-routine | 0.046*** | 0.030*** | 0.025*** | 0.008* | 0.011** |
| | (0.008) | (0.008) | (0.008) | (0.004) | (0.005) |
| Analytical non-routine x AI | -0.014 | -0.012 | -0.009 | -0.003 | -0.010** |
| | (0.010) | (0.008) | (0.009) | (0.004) | (0.004) |
| Interactive non-routine | 0.014** | 0.007 | 0.003 | -0.001 | 0.013** |
| | (0.007) | (0.008) | (0.009) | (0.004) | (0.006) |
| Interactive non-routine x AI | 0.001 | 0.006 | 0.006 | 0.000 | -0.011*** |
| | (0.009) | (0.008) | (0.008) | (0.003) | (0.003) |
| Manual routine | -0.031*** | -0.034*** | -0.031*** | -0.027*** | -0.012** |
| | (0.008) | (0.007) | (0.008) | (0.005) | (0.006) |
| Manual routine x AI | 0.017* | 0.015* | 0.015* | 0.006 | -0.002 |
| | (0.009) | (0.008) | (0.008) | (0.004) | (0.003) |
| Manual non-routine | -0.010 | -0.015* | -0.013 | -0.026*** | -0.016* |
| | (0.010) | (0.008) | (0.008) | (0.007) | (0.009) |
| Manual non-routine x AI | -0.017* (0.009) | -0.012 (0.008) | -0.012 (0.009) | -0.004 (0.005) | -0.011** (0.005) |
| Baseline Wage | Yes | Yes | Yes | Yes | Yes |
| Worker Controls | No | Yes | Yes | Yes | Yes |
| Firm Controls | No | Yes | Yes | Yes | Yes |
| occupation skill complexity | No | Yes | Yes | Yes | Yes |
| AKM Effects | No | No | Yes | Yes | Yes |
| Robot & Software Exposure | No | No | Yes | Yes | Yes |
| Region FEs | No | No | No | Yes | Yes |
| Industry FEs | No | No | No | Yes | Yes |
| Sales &Admin | No | No | No | Yes | No |
| 2-digit Occupation FEs | No | No | No | No | Yes |
| Observations | 1,444,158 | 1,444,134 | 1,345,353 | 1,345,353 | 1,345,353 |
| Adjusted R2 | 0.083 | 0.111 | 0.111 | 0.123 | 0.125 |

*Notes*: This table presents OLS estimates of the effects of AI exposure on the difference in log daily wage between 2012-2019 for females. Occupational AI, robot and software exposures are standardized exposure measures over the 5-digit kldb (2010) occupations, based on Webb (2020). Occupational task routine measures are standardized measures over the 3-digit kldb (2010) occupations, based on Dengler et al., (2014). The covariates included in the model are reported at the bottom of the table. Column 1 contains only occupational AI exposure controlling for baseline wages in 2012. Columns 2 and 3 include worker, occupation and firm controls (education, age, age squared, part-time work, AKM effects, occupational skill complexity and firm size) and robot and software exposure measures. Column 4 include region and industry (3 digit) fixed effects and controls for occupations belonging to either sales or administration. Column 5 includes 2-digit occupation fixed effects. Standard errors within parenthesis are clustered at the 5-digit occupation level. * p < 0.05, ** p < 0.01, *** p < 0.001.

*Source*: IEB (2012-2019) restricted to full-time female workers, liable to social security & Dengler et al. (2014).